\documentclass[12pt]{article}
\usepackage{epsf}
\def\la{\mathrel{\mathpalette\fun <}}
\def\ga{\mathrel{\mathpalette\fun >}}
\def\fun#1#2{\lower3.6pt\vbox{\baselineskip0pt\lineskip.9pt
\ialign{$\mathsurround=0pt#1\hfil
##\hfil$\crcr#2\crcr\sim\crcr}}}

\newcommand{\veR}{\mbox{\boldmath${\rm R}$}}

\newcommand{\veS}{\mbox{\boldmath${\rm S}$}}

\newcommand{\veE}{\mbox{\boldmath${\rm E}$}}
\newcommand{\ven}{\mbox{\boldmath${\rm n}$}}
\newcommand{\veF}{\mbox{\boldmath${\rm F}$}}
\newcommand{\ver}{\mbox{\boldmath${\rm r}$}}
\newcommand{\vex}{\mbox{\boldmath${\rm x}$}}
\newcommand{\vej}{\mbox{\boldmath${\rm j}$}}
\newcommand{\vnabla}{\mbox{\boldmath${\rm \nabla}$}}
\newcommand{\be}{\begin{equation}}
\newcommand{\ee}{\end{equation}}
\newcommand{\bc}{\begin{center}}
\newcommand{\ec}{\end{center}}

\newcommand{\mr}[1]{\mathrm{#1}}
\newcommand{\fr}[2]{\frac{#1}{#2}}
\newcommand{\lt}{\left}
\newcommand{\rt}{\right}

\newcommand{\rf}[1]{(\ref{#1})}
\newcommand{\lb}{\label}

\title{\bf Triangular and $Y$-shaped hadrons with static sources} 
\author{D.S. Kuzmenko and Yu.A. Simonov}
\date{\it Institute of Theoretical and Experimental
Physics,\\ 117218, B.Cheremushkinskaya 25, Moscow, Russia}

\begin{document}
\maketitle

\begin{abstract}
The structure of hadrons consisting of three static color sources 
in fundamental (baryons) or adjoint (three-gluon 
glueballs) representations is studied. The static potentials of 
glueballs as well as gluon field distributions in glueballs and 
baryons are calculated in the framework of field correlator 
method.  
\end{abstract}

1. The study of the gluon field  structure in hadrons is 
important for the deep understanding of the strong 
interaction physics. Interesting results on the baryon flux 
structure were obtained recently in the maximally abelian gauge on 
the lattice \cite{Ichie}. We demonstrate below how to perform 
calculations in gauge invariant way using the field correlator 
method relyed directly on QCD. 
 
The static potential in baryon was considered in detail
in \cite{1} at this conference, and 
we will extensively refer to this talk in what follows.
We start with the calculations of static 
potentials in three-gluon glueballs. This part of the talk
presents the results from \cite{2}. In the rest of the talk
 the gluon field distributions in hadrons are discussed on the 
base  of papers \cite{3}, \cite{4}, \cite{5}. 

In contrast to baryons, the gauge-invariant extended wave 
function of glueballs may have both $Y$-type and triangular 
structure. In the former case it may be written as
\be
 G^{(f)}_Y(x,y,z,Y)=f^{abc}g_a(x,Y) g_b(y,Y) g_c(z,Y)\quad \mbox{ 
or} 
\label{1}
 \ee
\be
 G^{(g)}_Y(x,y,z,Y)=g^{abc}g_a(x,Y) g_b(y,Y) g_c(z,Y).
\label{2}
 \ee
Here $g_a(x,Y)=g_a(x)\Phi^{ab}(x,Y)$ denotes the extended gluon
operator, $g_a(x)$ is the valence gluon operator and
$\Phi^{ab}(x,Y)=(\mr{P}\exp ig\int_Y^x A_\mu dz_\mu)^{ab}$ is
the parallel transporter or Schwinger line in the adjoint 
representation.
The coordinates  $x,y,z$ and $Y$ in \rf{1} apply to the valence
gluons and string junction positions respectively, $f^{abc}$ and 
$g^{abc}$ denote adjoint antisymmetric and symmetric symbols.

The wave function of the triangular glueball has the form
\be
G_\Delta(x,y,z) =G_\alpha^\beta (x) \Phi_\beta^\gamma(x,y)
G_\gamma^\delta(y) \Phi_\delta^\varepsilon (y,z) G_\varepsilon
^\rho(z) \Phi_\rho^\alpha(z,x),
\label{3}
\ee
where $G_\alpha^\beta (x)=g^a(x)(t^a)_\alpha^\beta$, $t$ is the
generator of $SU(3)$, and $\Phi_\alpha^\beta$ is the parallel
transporter in fundamental representation.

According to \rf{1}, \rf{2}, a Wilson loop of $Y$-type glueball
has the same structure as in the case of baryon, see eq. \rf{2} 
and Fig. 1 of \cite{1}. The difference is that the adjoint
transporters should be substituted for the fundamental  ones and 
symbols  $f^{abc}$ or $g^{abc}$ for 
$\epsilon_{\alpha\beta\gamma}$. 

A Wilson loop of triangular glueball, induced by \rf{3}
in the limit of large $N_c$, is shown in Fig. 1. The trajectories
of valence gluons are replaced here with two fundamental (quark)
trajectories and the Wilson loop is reduced to three 
disconnected rectangular meson loops.

The static potential in baryon calculated  in the 
field correlator method (MFC) is proportional to two-point
gluon field strength correlators  in fundamental representation,
see e.g. \rf{5}, \rf{6} from \cite{1}. Therefore it is 
proportional to the fundamental Casimir operator, $C_F=4/3$, and 
the ratio of $Y$-type glueball and baryon potentials equals
to the ratio of corresponding Casimir operators, 

\begin{figure}[!t]
\epsfxsize=8cm
\hspace*{4.35cm}
\epsfbox{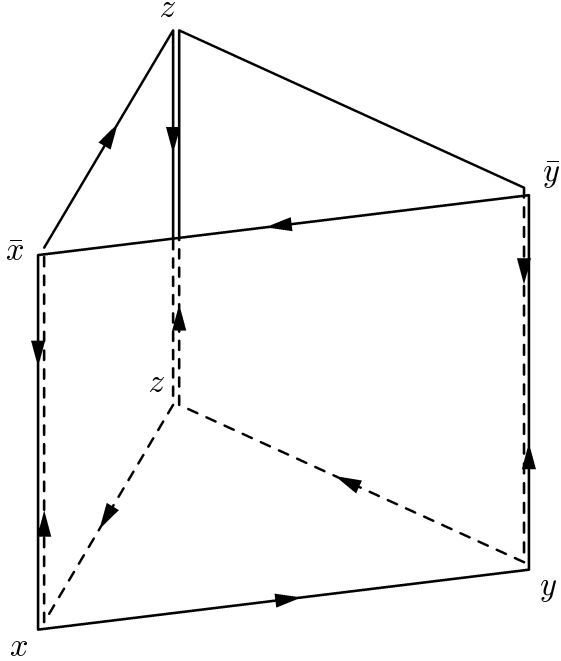}
\caption{$\Delta$-type Wilson loop}
\end{figure}

\be
\fr{V_{Y}^{(G)}}{V_Y^{(B)}}=\fr{C_A}{C_F}= \frac94,
\lb{4}
\ee
where $C_A=3$ is the adjoint Casimir operator and nondiagonal
(interference) terms are neglected.
The diagonal part of the baryon potential in equilateral triangle 
 has the form \cite{2}
\be
V_Y^{(B)}(R)=\fr{6\sigma}{\pi}\lt\{
R\int_0^{R/T_g}dx\, x K_1(x)-T_g\lt(2-\fr{R^2}{T_g^2}K_2\lt(
\fr{R}{T_g}\rt)\rt)\rt\}\equiv 3V^{(M)}(R),
\lb{5}
\ee
where $R$ is the distance from quarks to string junction,
$K_1,~K_2$ are McDonald functions, $\sigma$=0.18 GeV$^2$
and $T_g= 0.12\div 0.2$ fm are the string tension and the gluon 
field correlation length, and $V^{(M)}(R)$ is the static 
potential in meson. 

\begin{figure}[!t]
\epsfxsize=12cm
\hspace*{2.35cm}
\epsfbox{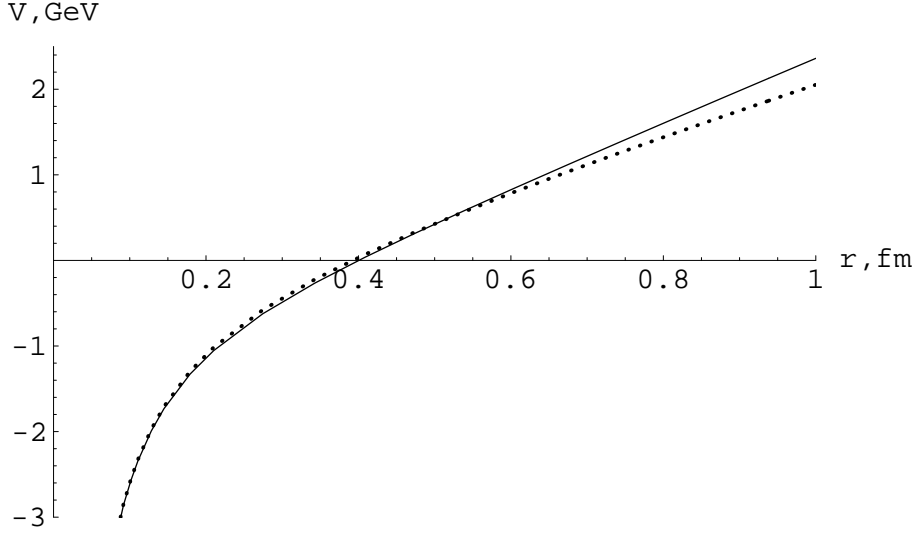}
\caption{Glueball potentials
$V_Y^{(G)}(r)+V^{\mathrm{pert}}_{\mathrm{(adj)}}(r)$  (solid 
curve) and  $V_\Delta^{(G)}(r)+ 
V^{\mathrm{pert}}_{\mathrm{(adj)}}(r)$ (dashed curve) in 
equilateral triangle with the quark separations $r$ for 
$\alpha_s=0.3$, $\sigma$=0.18 GeV$^2$ and $T_g=0.12$ fm. The 
nondiagonal terms are neglected.} 
\end{figure}

Now we turn to the triangular glueball, whose Wilson loop
according to Fig. 1 is the product of three meson loops. 
Therefore, the potential $V_\Delta^{(G)}$, which is the logarithm 
of the Wilson loop, is the sum of three meson potentials. In the 
case of equilateral triangle with the side $r=\sqrt{3}R$
\be
 V_\Delta^{(G)}(r)=3V^{(M)}(r).
\lb{6}
\ee
The perturbative potential for three-gluon glueballs
reads as
\be
V^{\mathrm{pert}}_{\mathrm{(adj)}}(r)=-\frac32\,
\frac{C_2(\mathrm{adj})\alpha_s}{r}.
\label{7}
\ee
The behavior of total potentials in $Y$-type and triangular 
glueballs are shown in Fig. 2. One can see that they are very
close up to distances $r\approx 0.6$ fm. Note that the 
interference terms are to change this picture. They will be
considered in subsequent publications. 

2. We proceed now to the study  of the gluon 
field structure using the connected probe \cite{6}, \cite{7}. The 
connected probe consists of the probe plaquette joined to the 
Wilson loop by parallel transporters and forms the frame with the 
current in four-dimentional euclidian space, see Fig. 3. When the 
plaquette size is small enough, the connected probe affords to 
calculate the color-integrated gluon field  in hadron, using the  
Wilson loop $\cal W^H$ of the latter,
\be
F^H_{\mu\nu}(x)= -ig \,\frac{\langle {\cal 
W^H}^\alpha_{\beta}(x_0) 
\Phi^{\beta}_{\gamma}(x_0,x)(F_{\mu\nu}^a(x)t^a)^{\gamma}_{\delta} 
(\Phi^+)^{\delta}_{\alpha}(x,x_0)\rangle }{\langle{\cal 
W^H}\rangle}. \lb{8}
\ee
The following  expression is valid in the bilocal approximation of 
MFC in the case of mesons \cite{4}, 
\be
F^M_{\mu\nu}(x)=\int_S d\sigma_{\rho\sigma}(x')\,
{\cal D}_{\rho\sigma,\mu\nu}(x'-x),
\lb{9}
\ee
where the integration is taken over the surface of Wilson loop $S$
and ${\cal D}$ denotes the gauge-invariant bilocal gluon field 
strenth correlator, see e.g. \rf{5} of \cite{1} for details.
Using the MFC parametrization of correlators \cite{1},
we calculate that the electric component of the $F^M$  has the 
form \cite{5}
\be
\veE^M(\vex,\veR)=\ven\, \fr{2\sigma}{\pi}
\int_0^{R/T_g} dl\, \lt| l\ven-\frac\vex{T_g}\rt|
K_1\lt(\lt| l\ven-\frac\vex{T_g}\rt|\rt),
\lb{10}
\ee
while the magnetic one is absent.

The field $\veE^M(\vex,\veR)$ \rf{10}, which we call the 
nonperturbative background gluon field in meson, 
is the force acting on the probe located at 
the point $\vex$ while the quark situated at zero and antiquark at 
the point $\veR$. It is related to the field acting on the quark
as follows.

  The force $\veF^M$ acting on the quark
in meson is defined by the static potential $V^{(M)}$ \rf{5},
\be
\veF^M(\ver)=-\vnabla V^{(M)}(\ver).
\lb{11}
\ee
\begin{figure}[!t]
\epsfxsize=10cm
\hspace*{3.35cm}
\epsfbox{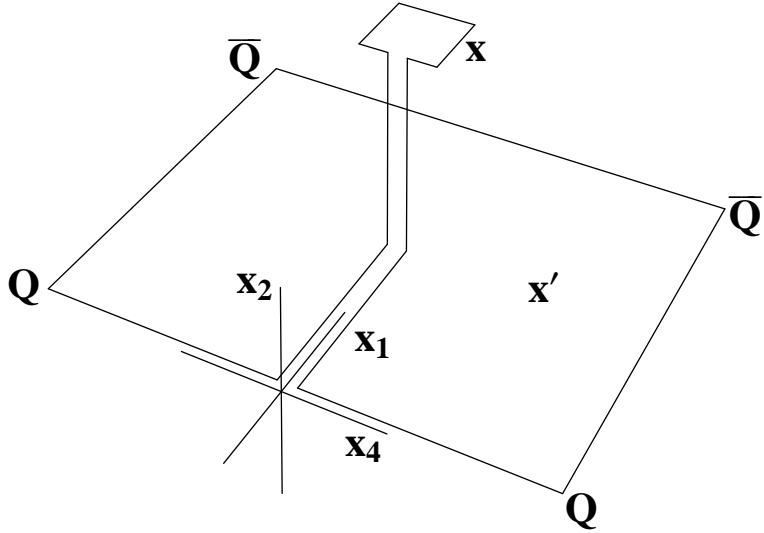}
\caption{A connected probe in the case of meson}
\end{figure}

One can verify using \rf{5}, \rf{10}, \rf{11} that the following 
relations are valid \cite{5}, 
\be
\veE^M(0,\veR)=\veE^M(\veR,\veR)=\veF^M(\veR),
\lb{12}
\ee
\be
\veE^M(\veR,2\veR)=\veE^M(0,\veR)+\veE^M(\veR,\veR)=
2\veF^M(\veR).
\lb{13}
\ee
The relation \rf{12} mean that when the locations of probe and 
quark (antiquark) coinside, the probe recombines with the latter,
and that is why it is affected by  the same force  $\veF^M$.
One can conclude from the relation \rf{13} that when the
probe is situated in the middle of two sources, it interacts
with both  the quark and antiquark, and the total field becomes
twice as big. If the point $\vex$ is located on the line
connecting the quark and antiquark, it is easy to check using 
\rf{5}, \rf{10} that the generalization of \rf{12},
\rf{13} has the form
\be
\veE^M(\vex,\veR)=\veF^M(\vex)-\veF^M(\vex-\veR).
\lb{14}
\ee

According to \rf{11}, the force 
$\veF^M(\veR)$ acting on the quark increases linearly
with the slope $2\sigma/(\pi T_g)$ at small distances $R\la T_g$
and saturates with the value $\sigma$ at $R\ga 0.7$ fm.

The field $\veE^M$ in the saturated regime acquires the universal
profile $E^{\mr{string}}$, which does not depend on the 
quark-antiquark separation,
\be
E^{\mr{string}}(\rho)=2\sigma\lt(1+\fr{\rho}{T_g}\rt)
\exp\lt(-\fr{\rho}{T_g}\rt),
\lb{15}                                                         
\ee                                                                                             
where $\rho$ is the distance from the quark-antiquark axis. 
We apply now the Gauss law in the form
\be
\int \veE\, d\veS=4\pi C_F\alpha_s
\lb{Gauss}
\ee
to the saturated field \rf{15} and get the parameter relation
\be
3\,\sigma T_g=C_F\alpha_s.
\lb{param}
\ee 
Taking the freezing value of the strong coupling 
 $\alpha_s\approx 0.4$ \cite{8} and the phenomenological value 
of the string tension $\sigma=0.18$ GeV$^2$, we determine from
\rf{param} the reasonable gluon correlation length value 
$T_g\approx 0.2$ fm  (see \cite{1}). 

\begin{figure}[!t]
\epsfxsize=14cm
\hspace*{2.35cm}
\epsfbox{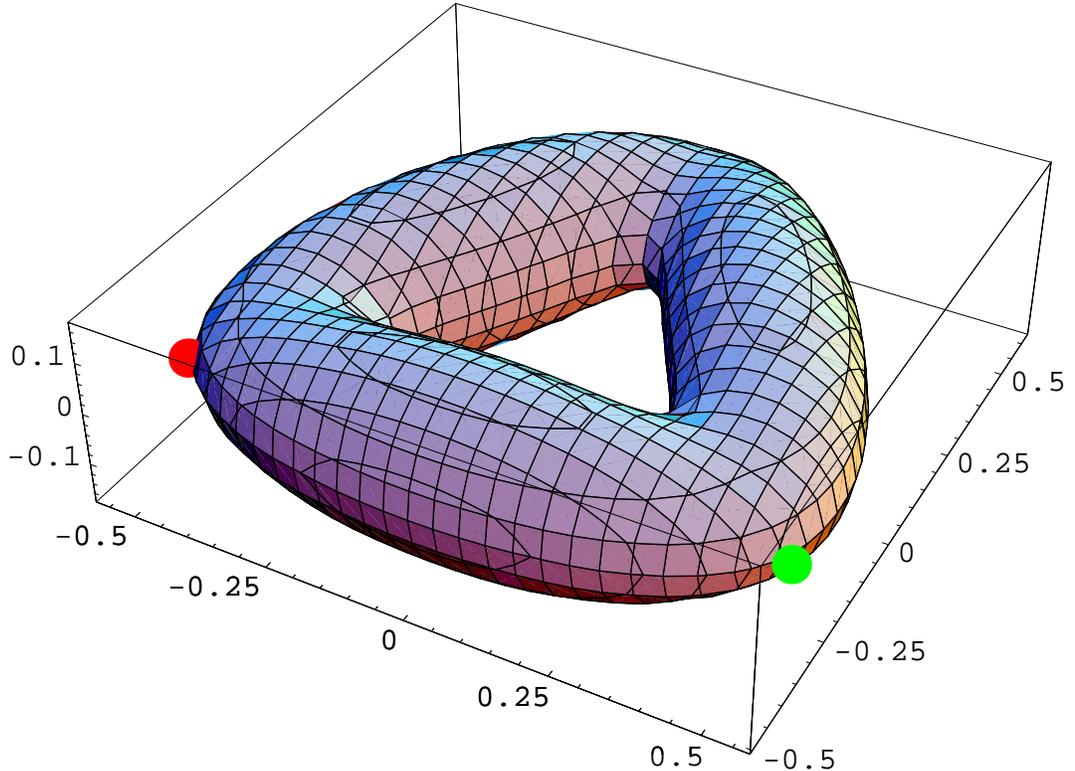}
\caption{The contour plot of the field in the $\Delta$-type 
three-gluon glueball. The valence gluon separations are 1 fm, 
$\sigma=0.18$ GeV$^2$, $T_g=0.12$ fm.}
\end{figure}

Using the Maxwell equation with  magnetic currents,  
\be
\mr{rot}\,\veE^M=\vej_{\mr{magn}},
\lb{Max}
\ee
we obtain that currents corresponding to the 
saturated field   \rf{15} form closed 
circles around the quark-antiquark axis, with the density
\be
j_{\mr{magn}}(\rho)=\fr{\partial E^{\mr{string}}(\rho)}{\partial 
\rho}=\fr{2\sigma \rho}{T_g^2}\exp\lt(-\fr{\rho}{T_g}\rt).
\lb{magn}
\ee 

One can verify using \rf{9}, \rf{Max} and nonabelian Bianchi 
identity that magnetic currents arise due to the three-point field 
strength correlator, which describes the emitting of the 
color-magnetic gluon field by the color-electric one.

The detailed study of 
these and other properties of the field $\veE^M$ will be given 
elsewhere. 

3. When the field distribution of the quark-antiquark pair is 
known, it is straightforward to calculate the field of triangular 
glueball (see Fig. 1). We join the connected probe to each of
quark-antiquark loops and arrive at the expression
\be
\veE_\Delta^{(G)}(\vex,\ver^{(1)},\ver^{(2)},\ver^{(3
)})=
\sum_{i=1}^3 \veE^M(\vex-\ver^{(i)}, \veR^{(i)}),
\lb{16}
\ee
where $\ver^{(i)}$ is the position of $i$-th valence gluon
and $\veR^{(i)}=\ver^{(i+1)}-\ver^{(i)}$. The surface defined by 
the condition $|\veE_\Delta^{(G)}(\vex)|=\sigma$ at gluon 
separations 1 fm is shown in Fig. 4. Values of parameters  
$\sigma=0.18$ GeV$^2$, $T_g=0.12$ fm are used. The surface shown 
in the figure goes through the valence gluon locations. Indeed, 
one can verify using \rf{16} that two forces of the value $\sigma$ 
having the angle $2\pi/3$ between them add to give the resulting 
force of the value $\sigma$. 

The same procedure can be applied to the system of arbitrary 
number of mesons to obtain the  field distributions 
in the leading order of $1/N_c$.   The next order corrections
can also be considered within the approach.

\begin{figure}[!t]
\epsfxsize=14cm
\hspace*{2.35cm}
\epsfbox{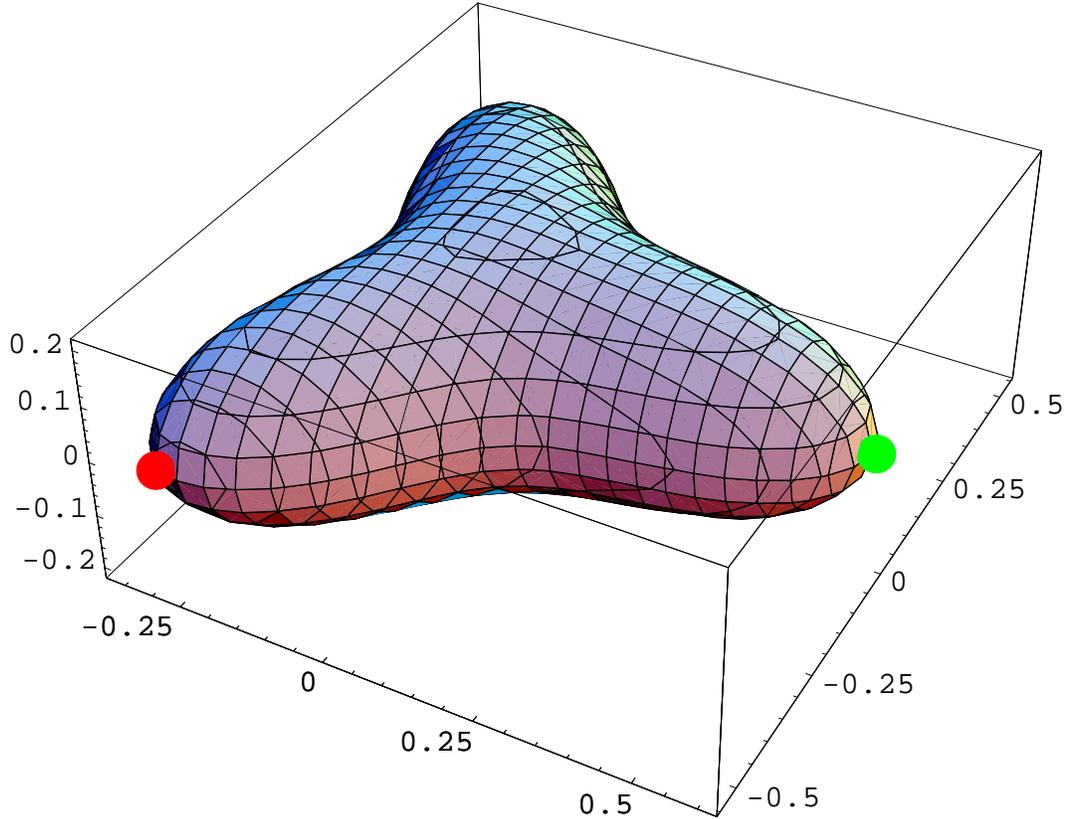}
\caption{The contour plot of the field in the baryon. The quark 
separations are 1 fm, $\sigma=0.18$ GeV$^2$, $T_g=0.12$ fm.} 
\end{figure} 

4. The problem of baryon (or $Y$-type glueball) field  
calculations is  more complicated. 
When we join the probe plaquette to the trajectory of the first 
quark and calculate the field distribution \rf{8} using the 
bilocal approximation of MFC, we arrive at the 
following relation for the electric field,  
\be
\veE^B_{(1)}(\vex,\veR^{(1)},\veR^{(2)},\veR^{(3)})=
\veE^M(\vex,\veR^{(1)})-\frac12\,
\veE^M(\vex,\veR^{(2)})-\frac12\,\veE^M(\vex,\veR^{(3)})
\lb{17}
\ee
(the magnetic field is absent just as in the meson case).
The vector $\veR^{(i)}$ here is directed from the string junction 
to the $i$-th quark, $\veE^M$ is defined in \rf{10}. 
When we join the probe to the second and third quark trajectories, 
we will get analogous formulas with the corresponding index
transpositions.

 To  obtain symmetric field distributions,   
we are to sum squares  $(\veE^B_{(i)})^2$,
\be
(\veE^{(B)})^2=\frac23\, \lt((\veE^B_{(1)})^2
+(\veE^B_{(2)})^2+(\veE^B_{(3)})^2\rt),
\lb{18}
\ee
The surface given by the condition
 $|\veE^{(B)}(\vex)|=\sigma$ at quark separations 1 fm is shown in 
Fig. 4. It goes through the quark positions and has small 
convexity near the string junction. 

The ratio of the energy density at the 
string junction position, $w_{\mr{sj}}$, to the density in the 
middle of the string with the saturated profile, 
$w_{\mr{string}}$, has to be equal to the corresponding field 
squares and according to \rf{18}, \rf{15} 
\be
\fr{w_{\mr{sj}}}{w_{\mr{string}}}=\fr{(\veE^{(B)}(0))^2}{
(\veE^{\mr{string}}(0))^2}=\fr98.
\lb{19}
\ee

To conclude,  hadron static potentials  
and the gluon  field distributions,  which ensure  the confinement 
of color sources, were  calculated using the field correlator 
method. It is stressed that the mechanism of confinement is  
the emitting process of the color-magnetic gluon field by  the
color-electric one in the nonperturbative vacuum.
The static potentials of three-gluon glueballs were 
 reduced in the first approximation to the 
appropriate sum of quark-antiquark potentials. 
The nonperturbative gluon field  induced by the 
static quark-antiquark pair was calculated using the connected 
probe, and its classical abelian properties were considered. 
Triangular gluon field distributions were calculated 
for glueballs and  $Y$-type ones for baryons. 

An extension of the method to the arbitrary number of 
colors is straightforward. It is also applicable to  
the  nuclear structure study. 

This work has been supported by 
 RFBR grants 00-02-17836,  00-15-96786, and INTAS 00-00110, 
00-00366.

\end{document}